\documentstyle[preprint,tighten,aps]{revtex}

\begin{document} 

\def \re#1 {(\ref{#1})}
\def \Q {{\cal  Q}}
\def \k {\kappa}

\def \ha{{\textstyle{1\over 2}}}

\def \D {\Delta}
\def \a {\alpha}
\def \b {\beta}
\def\g {\gamma}
\def\r {\rho}
\def \s {\sigma}
\def \p {\phi}
\def \m {\mu}
\def \n {\nu}
\def \vp {\varphi }
\def \l {\lambda}
\def \t {\theta}
\def \td{\tilde }
\def \d {\delta}
\def \ci {\cite}
\def \la {\label}
\def \sm {$\s$-model }
\def \foot {\footnote }
\def \P {{\cal P}}
\def \o {\omega}
\def \inv {^{-1}}
\def \ov {\over }
\def \four{{\textstyle{1\over 4}}}
\def \fourth{{{1\over 4}}}
\def \foot{\footnote}
\def\be{\begin{equation}}
\def \ee{\end{equation}}
\def\bea{\begin{eqnarray}}
\def\eea{\end{eqnarray}}
\def \sh {\ {\rm sinh}}
\def \ch {\ {\rm cosh}}
\def \cosh {\ {\rm cosh}}
\def \sinh {\ {\rm sinh}}
\def \cosh {\ {\rm cosh}}
\def \tanh {\ {\rm tanh}}
\def \th {\ {\rm tanh}}


\draft
\preprint{DAMTP/R/96/27, \  Imperial/TP/95-96/52, \  hep-th/9606033} 
\date{June 1996}
\title{ Non-Extreme Black 
Holes from Non-Extreme Intersecting M-branes}
\author{Mirjam Cveti\v c$^1$
\thanks{On sabbatic leave from University of Pennsylvania. 
E-mail address: cvetic@cvetic.hep.upenn.edu}
and A.A. Tseytlin$^2$
\thanks{Also at Lebedev Physics Institute, Moscow.\ 
E-mail address: tseytlin@ic.ac.uk}}
\address{$^1$ 
Department of Applied Mathematics and Theoretical Physics,\\University 
of Cambridge, Cambridge CB3 9EW, U.K.
\\and\\
$^2$ Theoretical Physics Group, Blackett Laboratory,\\ Imperial 
College, London SW7 2BZ, U.K.}
\maketitle \begin{abstract}
{We present   non-extreme generalisations of intersecting p-brane solutions  
of 
eleven-dimensional supergravity which upon toroidal compactification 
reduce to non-extreme static black holes
 in dimensions $D=4$,
$D=5$ and $6\le D\le 9$,   parameterized 
by four, three and two charges, respectively.  
The $D=4$ black holes are obtained  either from a non-extreme configuration of 
three  intersecting 
five-branes with a  boost  along the common string or from non-extreme  
intersecting system  of  two   two-branes and two 
five-branes. 
The $D=5$ black holes arise 
 from three intersecting two-branes or from a system of  intersecting  
two-brane and 
five-brane with a  boost along the common string. 
Five-brane and two-brane with a boost along one direction reduce 
to black holes in $D=6$ and $D=9$, respectively,  while 
$D=7$  black hole 
can be interpreted in terms of   non-extreme configuration of 
two   intersecting  two-branes. 
 We discuss the expressions for the 
corresponding masses and entropies.

}
 \end{abstract} 

\section{Introduction}

Recently,  black holes in string theory  have become a subject of
intensive research,  due, in part,  to the fact that  microscopic
properties, e.g.,  the statistical origin of the   entropy
of certain black holes  can be addressed either by  using  
  a  conformal field theory description  of  the NS-NS  
  backgrounds  \cite{Sen,LW,CTII,TMpl,LWII,CYNear,TLast}
or  a  $D$-brane  representation  of U-dual  R-R backgrounds 
\cite{SV,CM,HStrom,BMPV,MSTR,JKM,MSS,Horowitz}.

By now,  the classical solutions for the  BPS-saturated as well as  
non-extreme   static and  rotating black holes
 of $N=4,8$ supersymmetric  superstring vacua
 are well
understood.   In particular,  the 
explicit form of the  generating solution for  general  rotating
black holes  has been obtained.  The generating
solution 
is specified by the  canonical choice of the asymptotic values
of the scalar fields, the ADM  mass,   $[{{D-1}\over 2}]$ components of angular
momentum
and  by five, three and two charges, in  dimensions $D=4$, $D=5$ and $6\le 
D\le
9$, respectively.\footnote{This program has been completed in
 $D=5$\cite{CY5r}
and $D\ge 6$\cite{CYNear,Llatas};  however,
 in $D=4$ only  the five-charge  static  generating solution
\cite{CY4s} (see also \cite{JPM})
 and the four-charge rotating solutions \cite{CY4r} were obtained.  The
BPS-saturated generating solutions  were obtained  earlier:
  in $D=4$ in  \cite{CY,CTII}  and in  $D=5$ in  \cite{TMpl}. This work was
preceded by a number of  papers, where special examples of such
solutions were obtained (for a recent review, see
 \cite{Horowitz}).}

The explicit from of the generating solutions was   first
determined in the case of   toroidally compactified heterotic string 
($N=4$ superstring vacua) or  in the
NS-NS   sector of the  toroidally
 compactified type
IIA superstring.\foot{The most general black hole  configuration
 is   obtained by applying to   the generating solution  a subset of
$T$- and $S$-duality  transformations (for $N=4$ superstring vacua)
\cite{Sen} 
or   $U$-duality
transformations (for $N=8$ superstring vacua) \cite{CH}, which do not affect 
the canonical choice of the
asymptotic values of the scalar fields. These transformations do
not change  
the $D$-dimensional Einstein-frame metric,  and thus  the  metric 
of a general black hole in this class is thus  fully specified by the
parameters of the generating solution. 
A solution with arbitrary
asymptotic values of the scalar fields is found  by appropriate
rescalings of   
the physical parameters, i.e. of the ADM mass, the
angular momenta and the charges,   by the asymptotic values of the
scalar fields.}
     By applying  $U$-duality transformations such
solutions   are mapped  onto backgrounds 
with  R-R charges which 
have an  interpretation in terms of $D$-brane configurations.
These   BPS-saturated  black holes  have    regular horizons 
and finite semiclassical Bekenstein-Hawking (BH) entropy  in 
dimensions $D=4$ \cite{KLOPV,CY,CT,CTII} and $D=5$ \cite{SV,TMpl}. 
In $D\ge 6$ the  BPS-saturated  axi-symmetric solutions have 
singular horizons  and zero  BH entropy \cite{Peet,HSen,KT,CYNear}.

Entropy  of certain non-extreme or near-extreme  black holes 
was discussed  (both in the static and  rotating cases)  in $D=5$ 
\cite{HStrom,HMS,BLMPSV} and $D=4$ \cite{HLM}  and for rotating
black holes with NS-NS (electric) charges in $6\le D \le 9$ \cite{CYNear}.

A  unifying treatment of string-theory 
 black hole properties  may
arise by identifying  such black holes as (toroidally) compactified
configurations of intersecting  two-branes  and five-branes of
eleven-dimensional M-theory \ci{DVV,KTT}. 
The $D=10$  backgrounds  with  NS-NS  and R-R charges 
appear on an equal footing when viewed from  eleven dimensions.
 A discussion of  intersections  of certain
BPS-saturated  M-branes along with a proposal for intersection rules was
first given in \cite{PT}. A generalization to a number of
different harmonic functions specifying  intersecting
BPS-saturated M-branes  which  led to  a better understanding
 of these solutions and a   construction of 
new intersecting p-brane  solutions  in $D \leq 11$ was presented 
 in  \cite{TM} (see
also  related work \cite{KTT,CS,BE,GA,BL,PO}).
 Specific  configurations of that type reduce to the 
BPS-saturated black holes  with regular horizons 
in $D=5$ \cite{TM}  and $D=4$  \cite{KTT} 
 whose properties are determined by three and four charges (or 
harmonic functions), respectively.

 The purpose of the present 
paper is  to  relate   the {\it non-extreme} static black holes  
to {\it non-extreme } versions  of  intersecting  
 M-brane solutions of \cite{TM,KTT}. 
This approach  may shed light on the 
structure of  non-extreme black holes from the point of view of 
M-theory, and, in particular, clarify 
 the  origin of  their  BH  entropy.
Our interpretation of non-extreme black holes 
as non-extreme intersecting M-branes (or p-branes in $D=10$) 
does not seem  to  be related to 
the  ``brane-antibrane'' picture  suggested in \cite{CM,HMS,HLM,KKK}.

As we shall discuss in Section II, there exists a
 procedure allowing 
one to construct a non-extreme version of a given
BPS-saturated intersecting M-brane solution
which generalises the approach of \cite{TM}.
The  resulting eleven-dimensional  metric and the four-form field strength 
 depend  on  the  ``non-extremality'' parameter,  representing  a  deviation
from the BPS-saturated limit, and  the  ``boosts'', specifying charges
 of the configuration.
  Upon dimensional reduction  these
parameters  determine the  ADM  mass and the charges
 of non-extreme static  black holes in $4\le D 
\le
9$.\footnote{The non-extremality parameter $\mu$ is proportional to  the  
ADM mass of the neutral 
(Schwarzschild) black hole and the ``boosts'' 
can be interpreted  as parameters 
of symmetry transformations of the effective
lower-dimensional
 action  which generate   the charged solutions  when 
applied to the neutral black hole solution.} 

 In Sections III, IV, and V we  shall consider examples of non-extreme
configurations of intersecting    M-branes
and relate them, via dimensional reduction along internal 
M-brane directions,  to non-extreme black holes in dimensions $D=4$,
$D=5$ and $6\le D\le 9$,  respectively.
In Section  VI we shall present the general expressions for the mass
and BH entropy of these solutions
and comment on  some of their consequences.

\section{Non-Extreme 
Intersecting M-brane Solutions}
The aim  is to generalise the extreme  supersymmetric   (BPS-saturated)
intersecting M-brane
solutions of \cite{TM,KTT}   
to  the non-extreme case.
Even though non-extreme solutions are no longer  supersymmetric, 
it turns out that they can be constructed as a 
``deformation'' of  extreme solutions, parameterised by several
one-center harmonic functions $H_i$, one for each constituent M-brane, 
and the Schwarzschild solution,  
 parameterised by  the  function $f(r)=1-\mu/r^{D-3}$.\foot{Similar  ``product 
structure''  was found previously for   non-extreme versions of isotropic
extreme   black  p-brane  solutions  in \cite{DLP,KT}.} 
 Here $D$ is the dimension of
 the space-time  transverse  to  the configuration, and 
the  ``non-extremality'' parameter $\mu$ specifies a deviation from the
BPS-saturated limit.
The same  type of construction  applies also to intersecting p-brane solutions 
in  ten dimensions.

It should be stressed  that the  solutions presented in this paper
should not be interpreted as intersections of non-extreme M-branes,
even though they reduce to single non-extreme M-brane solutions when all other 
charge parameters are set equal to zero. Each of non-extreme M-branes is 
parameterised,  in general, by independent 
masses and charges while the  non-extreme version of intersecting M-brane 
solutions has  only one common  mass parameter -- the 
non-extremality parameter  $\mu$.  Such configurations should be viewed  as 
non-extreme ``bound-state'' configurations.  
Note also that   the non-extreme solutions below  are  one-parameter 
``deformations'' 
of 
a  special 
type of supersymmetric  solutions  \cite{TM,KTT},
for which all of the harmonic functions are chosen to have a simple
spherically symmetric one-center form.
The non-extreme version of multi-center solutions (corresponding to individual p-branes `placed' at  different points in transverse space)
are expected to be unstable, 
i.e. described by time-dependent background fields. On the contrary, the non-extreme versions of single-center solutions discussed below are static.

One way of understanding  why the structure of the non-extreme solutions is 
similar to  that of extreme ones
is  based on first doing  a dimensional reduction  to $D=10$, applying 
$T$-duality transformation  and then lifting the solution back to $D=11$.
Since a non-extreme solution has the same number of isometries as the extreme 
one, 
it can be ``generated'' by starting from  the
 Schwarzschild background
instead of the flat space one  by  $T$-duality 
considerations similar to the ones 
used in the extreme case in   \cite{BE,GA}. 
Alternatively, one may start with extreme solution 
and consider its deformation  caused by turning  on the 
 non-extremality parameter  $\mu$.

A simple algorithm  which leads to 
non-extreme version of a given extreme solution
(which indeed can be checked to satisfy the eleven-dimensional 
supergravity equations of motion and also  corresponds  upon dimensional
reduction to known non-extreme black hole solutions) 
consists of the following steps:

(1) Make the  following  replacements  in the  $D$-dimensional 
{\it transverse}  space-time  part
of the metric:
\begin{equation}
dt^2  \to   f(r) dt^2 \ , \ \ \ \ 
dx_n dx_n \to f^{-1} (r) dr^2 + r^2 d \Omega^2_{D-2}\ , \ \ \ 
\ f(r) = 1 - {\mu\over  r^{D-3}} \ , 
\label{ii}
\end{equation}
and also use the special one-center  form of the harmonic  functions,
\begin{equation}
H_i = 1 + {\Q_i\over r^{D-3}} \ , \ \ \ \ \  \ \ \ \Q_i= \mu\sinh^2 \delta_i \ 
, 
\label{uu}
\end{equation}
for the constituent   two-branes,
and 
\begin{equation}
H_i = 1 + {\P_i\over r^{D-3}} \ , \ \ \ \ \  \ \ \ \P_i= \mu\sinh^2 \gamma_i \ 
, 
\label{uup}
\end{equation}
for the constituent   five-branes.

(2) In the expression for the field strength  ${\cal F}_4$ of the
 three-form field make the following 
replacements: 
\begin{equation}
 H'_i \ \ \to \ \ H'_i = 1 + {Q_i\over r^{D-3} + \Q_i - Q_i } 
  =\big[ 1- { Q_i \ov r^{D-3}} H_i\inv\big]\inv \  , \ \  \ \ \ \ \ 
 Q_i =\mu\sinh\delta_i\cosh\delta_i\ , 
\label{ll}\end{equation}
in the ``electric''  (two-brane) part, 
and  
\begin{equation}
 H_i \ \ \to \ \ H'_i = 1 + {P_i\over r^{D-3}  }  \  , \ \ \ \ \ \ \ \ \ 
 P_i =\mu\sinh\gamma_i\cosh\gamma_i\ , 
\label{llp}\end{equation}
in the ``magnetic''  (five-brane) part.
Here $Q_i$   and $P_i$ 
are  the respective  ``electric'' and ``magnetic''  charges
of the configuration. In the  extreme   limit 
$\mu \to 0 ,\  \delta_i\to \infty$, and $\gamma_i\to  \infty$,
 while the charges $Q_i$ and $P_i$
are  kept fixed. In
this case $\Q_i=Q_i$ and  $\P_i =P_i$, so that  $H'_i=H_i$.
The form of  ${\cal F}_4$ and the actual value of its ``magnetic''
part does not   change compared to the extreme limit.

(3) In the case when the extreme solution has a  null isometry,
i.e. intersecting branes have  a common string along some  direction $y$, 
one can add momentum along $y$ by  applying  the coordinate 
transformation 
 \begin{equation}
t'= \cosh \beta \ t - \sinh \beta\ y\ ,  \ \ \ \ \ 
\ y'= - \sinh \b\ t + \cosh \beta\ y \ , 
\label{iou}
\end{equation}
to the   non-extreme background obtained 
according to the above two steps. Then 
$$
 - f(r) dt^2 + dy^2\  \to  \  - f(r) dt'^2 + dy'^2
= - dt^2 + dy^2 +
 {\mu\over r^{D-3}} (\cosh \beta\ dt - \sinh \beta\ dy)^2
$$ 
\begin{equation}
 = \   - K^{-1} (r) f(r) dt^2 + K(r) \widehat{dy}^2 \ ,  \ \ 
\ \ \ \ 
\ \widehat {dy}\equiv  dy  + [{K'}\inv (r) -1] dt\ , 
\label{yy}
\end{equation}
\begin{equation}
K= 1 + {\tilde \Q \over r^{D-3}} \ , \ \ \ \  
 {K'}\inv  = 1- { \td Q \ov r^{D-3}}  K \inv \ , 
\ \ \ \   \ \  \tilde \Q = \mu \sinh^2 \beta \ , \ \ 
\tilde Q = \mu  \sinh \beta \cosh \beta\ , 
\label{tt}
\end{equation}
where the 
 boost  $\b$  is related to the new electric 
charge   parameter $\tilde Q$, i.e. momentum along direction $y$.
In the extreme limit 
$\mu\to 0, \ \beta\to \infty$, the charge $Q$ is held fixed, $K={K'}$ 
and  thus  this part of the metric  (\ref{yy}) becomes
$ dudv + ({K}-1)du^2,$ where   $ v,u=y\pm  t$.

Below we shall  illustrate this algorithm on several examples.
Let us start with basic non-extreme M-brane
solutions found in \cite{Guven}.
The two-brane  background has the form\foot{We shall follow
\cite{TM} and use the notation  $T$ and $F$ for the 
inverse powers of the  harmonic functions corresponding 
to the two-brane and five-brane, respectively.}
\begin{eqnarray}
d s^2_{11} = T^{-1/3} (r) 
 \bigg( T(r) [- f (r)  dt^2  + dy^2_1 +  dy^2_2]
      + f^{-1}(r) dr^2+ r^2 d\Omega^2_{7}  \bigg),
\label{mem}
\end{eqnarray}
\begin{eqnarray}
{\cal F}_4 =  -3dt\wedge dT'\wedge  dy_1\wedge d y_2 \  , 
\label{fef}
\end{eqnarray}
where
\begin{equation}
 f= 1 - { \m\over r^6} \ , \ \ \ T^{-1}  = H= 1 + {\Q \over r^6}\ , \ \ \   
{T'} = {H'}\inv = 1 - {Q \over 
r^6} T \  , 
 \label{hhh}
\end{equation}
$$
\Q= \mu \sinh^2 \delta\ , \ \ \ \ Q=\mu\sinh\delta \cosh\delta\  .$$ 
Again, the extreme solution is obtained by setting $f=1, \ \Q=Q, \
T=T'$.\footnote{ 
  The relation  of our notation to the notation 
 used in  \cite{Guven}
is the following. The radial coordinate  of the $D$-dimensional  transverse
space-time in  \cite{Guven}
 is 
$\hat r^{D-3} = r^{D-3} + \Q = r^{D-3} H (r) ,
 \  \Q \equiv  r^{D-3}_- $  and   
$ 
 \Delta_- (\hat r) =  H^{-1}(r) , \ 
\Delta_+ (\hat r) =  H^{-1}(r) f(r),$  i.e. 
\ $\Delta_\pm (\hat r) =  1- {r^{D-3}_\pm/\hat r^{D-3}}$. Here  again
\   $r^{D-3}_+ = \mu \cosh^2\delta, \ 
r^{D-3}_- = \mu \sinh^2\delta$, 
 with $r^{D-3}_\pm \to  Q$ in the extreme limit. Note also 
that  in terms of $\hat r$,  $T'$ has the form 
$T'={H'}\inv = 1 - Q/\hat r^{D-3}$ .}

The five-brane  solution is
\begin{eqnarray}
d s^2_{11} = F^{-2/3} (r) 
 \bigg( F(r) [- f (r)  dt^2  + dy^2_1 + ...+  dy^2_5  ]
      + f^{-1}(r) dr^2+ r^2 d\Omega^2_{4}  \bigg)\ ,
\label{fiv}
\end{eqnarray}
\begin{eqnarray}
{\cal F}_4 =  3 *dF'^{-1} \  , 
\label{fif}
\end{eqnarray}
where the   dual form   is defined with respect to the flat transverse space. 
The  parameters  of the five-brane solution are ``magnetic''
 analogues of the ``electric'' parameters $\delta,\ \Q,\ Q$  of the
  two-brane solution  (\ref{hhh}) 
 and   are  denoted by $\gamma,\ \P\ ,P$, i.e. 
\begin{equation}
  f= 1 - { \mu \over r^3} \ ,\ \ \ F^{-1}=H = 1 + {\P \over r^3}\ , \
\ \  {F'}\inv  = {H'} = 1 + {P \over
r^3} \  , 
\label{fiif}
\end{equation}
$$   
\P= \mu \sinh^2 \gamma\ , \ \ \ \  P=\mu\sinh\gamma\cosh\gamma \ . $$
The two other non-extreme solutions found in \cite{Guven}
correspond to two and  three intersecting two-branes with equal values of
parameters $\delta_i=\delta$. 

A generalisation to the case of 
different  parameters $\delta_i$  can be easily found using the above 
algorithm.
For example, the non-extreme version of $2\bot 2$ configuration, i.e. two 
two-branes intersecting at a point,    is thus  given by\footnote{For the sake
of simplicity, in what
 follows  we  often  do 
not indicate explicitly  the  argument $r$  
(the radial transverse coordinate)  of 
 the  functions $T_i, F_i$ and $f$.}
\begin{eqnarray}
d s^2_{11} = (T_1T_2)^{-1/3}
 \bigg[ - T_1T_2 f   dt^2  +  T_1 (dy^2_1 +  dy^2_2) 
+ T_2 (dy^2_3 +  dy^2_4)
      + f^{-1} dr^2+ r^2 d\Omega^2_{5}  \bigg]\ ,
\label{mema}
\end{eqnarray}
\begin{eqnarray}
{\cal F}_4 =  -3dt\wedge (dT'_1\wedge  dy_1\wedge d y_2 
+   dT'_2 \wedge  dy_3\wedge d y_4 ) \  , 
\label{fefu}
\end{eqnarray}
where ($i=1,2$) 
\be
 f= 1 - { \m\over r^4} \ , \ \ \ \ T^{-1}_i = 1 + {\Q_i \over r^4}\ ,
\ \ \ \   T'_i= 1 - {Q_i \over r^4} T_i \  
, 
\ \label{popq}
\end{equation}
$$\Q_i= \m \sinh^2 \d_i\ , \ \ \  \ \ \  Q_i=\mu\sinh\delta_i\cosh\delta_i  \ .
$$
For $T_1=T_2$, $T_1'=T_2'$ this reduces to  the anisotropic
 four-brane solution 
of 
\cite{Guven}. The non-extreme version of  extreme $2\bot2\bot 2$
configuration (three two-branes intersecting at a point)
\cite{PT,TM}  has a similar form and will be discussed  below in Section IV.

 In the following Sections  we shall  construct the
non-extreme configurations  of  intersecting M-branes which   in $D\le 
9$ reproduce the generating solutions for  non-extreme static  black hole 
backgrounds. 
They will be built  in terms of 
basic M-branes according to the above algorithm.
As in \cite{PT,TM} we shall  consider only 
intersections which  in the extreme limit  preserve supersymmetry: 
   two two-branes can intersect at a 
point, 
two five-branes can intersect at a three-brane (with three-branes 
allowed to intersect over a string) and  five-brane and  two-brane 
can intersect at a string.

\section{$D=4$ Non-Extreme Black holes}\label{4D}
Four-dimensional  black holes with  
four independent charges\footnote{Note that the  
generating solution for the most general four-dimensional static black
holes of $N=4,8$ superstring vacua is specified by {\it five}
independent charges.}
can be obtained  upon toroidal compactification  from  {  two}  different
intersecting M-brane configurations \cite{KTT},  $2\bot 2\bot 5\bot 5$ 
and ``boosted'' $5\bot 5 \bot 5$.
 While the two  resulting black hole backgrounds 
are related   by   four-dimensional
$U$-duality,   the underlying
 intersecting M-brane solutions  should   be related by a symmetry
transformation of the M-theory. Such a symmetry transformation 
 should be obtained as  a combination   of $T$-duality and $SL(2,Z)$ symmetry 
of 
 the $D=10$ type IIB theory ``lifted'' to $D=11$.

\subsection{Intersection of two Two-Branes and two Five-Branes}

The first  of the above eleven-dimensional configurations
   corresponds to the  two   two-branes intersecting at a 
point and two  five-branes intersecting at a three-brane,  with  each of 
the two-branes intersecting  with each of the five-branes at a string.
The non-extreme version of 
 the BPS-saturated solution  found  in \cite{KTT}
is given by
$$
d s^2_{11} = (T_1 T_2)^{-1/3} (F_1 F_2 )^{-2/3}
 \bigg[- T_1 T_2 F_1 F_2 f dt^2  + F_1  ( T_1 dy^2_1 + T_2 dy^2_3)
    + F_2  ( T_1 dy^2_2+T_2 dy^2_4) 
$$
\be
 + \   F_1 F_2 ( dy^2_5 + dy^2_6
      + dy^2_7) + f^{-1}dr^2+ r^2d\Omega_2^2  \bigg]\ ,
\label{4d11d}
\end{equation}
$$
{\cal F}_4
 = -3dt\wedge (dT'_1\wedge  dy_1\wedge d y_2 +
  dT_2'\wedge  dy_3\wedge d y_4     )
$$
\begin{eqnarray}
+\  3( *dF'^{-1}_1 \wedge dy_2 \wedge dy_4 +  *dF'^{-1}_2
\wedge dy_1 \wedge dy_3 ) \  . 
\label{4dff}
\end{eqnarray}
The  
 coordinates $y_1, ... , y_7$ describe the toroidally
compactified  directions.  The 
 function $f$, parameterising a deviation from the 
extremality, and  functions $T_i,T'_i$ and $F_i,F'_i$,  specifying the
 (non-extreme)  two-brane  and 
five-brane configurations  depend on the
 radial coordinate $r$ of $(1+3)$-dimensional (transverse) space-time, 
\begin{equation}
f=1-{{\m}\over r}\ ,
\ \ \ \ \ 
T_{i}^{-1}=1+{\Q_i \over r}\ ,\ \ \  \ \ {T'_{i}}=1-{Q_i \over r}T_i \ , 
\label{4dtf}
\end{equation}
$$ \Q_i = \m \sinh^2\delta_i  \ , \ \ \ \ Q_i = \m \sinh \delta_i \cosh 
\delta_i  \ , \ \ \ i=1,2 \ , $$
\begin{equation}
F_{i}^{-1}=1+{\P_i \over r}\ ,\ \ \  \ \ {F'}_{i}\inv =1+{P_i \over r} \ , 
\label{fff}
\end{equation}
$$ \P_i = \m \sinh^2\gamma_i \ , \ \ \ \ P_i = \m \sinh \gamma_i \cosh
 \gamma_i \ , \ \ \  i=1,2
 \ . $$
In the extreme limit  $\mu\rightarrow 0$, 
$\delta_{i}\rightarrow \infty$ and  $ \gamma_i \to \infty$,  while 
the charges  $Q_i$ and $P_i$ are held fixed.
Again, in this limit $f= 1$, \ $T_i=T_i'$, \ $F_i=F'_i$.

The nine-area of the regular  outer horizon $r=\mu$ 
of the anisotropic seven-brane  metric 
 (\ref{4d11d})  is
\begin{equation}
A_9 =  4\pi 
L^7[ r^2 
(T_1 T_2 F_1 F_2)^{-1/2}]_{r=\m} = 4\pi L^7\m^2 
\cosh\delta_1  \cosh\delta_2 \cosh\g_1 \cosh\g_2 \    ,
\label{4darea}
\end{equation}
where the internal directions   $y_1,...
,y_7$ are assumed to have periods  $L$. 
In the BPS-saturated limit the area reduces to 
\begin{equation}
(A_{9})_{BPS}=4\pi L^7\sqrt{Q_1Q_2P_1P_2}\ .
\label{4dareabps}
\end{equation}
Upon toroidal compactification to four dimensions  one finds the following  
 Einstein-frame metric 
\begin{equation}
ds^2_4=-\lambda(r) f(r) dt^2+\lambda^{-1}(r)\big[f^{-1}(r)dr^2+
r^2d\Omega_2^2\big
]\  ,
\label{4dst}
\end{equation}
where  
\begin{eqnarray}
\lambda (r)= (T_1T_2F_1F_2)^{1/2} 
= {{r^2}\over [
(r+\Q_{1}) 
(r+\Q_{2}) 
(r+\P_1) 
(r+\P_2)]^{1/2} }\  .
\label{4dlambda} \end{eqnarray}
In the BPS  limit $f=1$ and  $\Q_i\to Q_i, \ \P_i \to P_i$.

The four-dimensional metric  (\ref{4dst})
 is precisely the one of the 
non-extreme four-dimensional black hole with two electric
and two  magnetic charges found in  
 \cite{CYNex}.

\subsection{Intersection of three Five-Branes with a Boost}

 The second relevant configuration \cite{KTT} is that of  three  
  five-branes, each pair 
intersecting at a three-brane, with an extra boost along a  string common to 
three three-branes.
 The corresponding  non-extreme background has the form 
$$
d s^2_{11}=  (F_1 F_2 F_3)^{-2/3}
  \bigg[ F_1 F_2 F_3 (- K^{-1} f dt^2 + K \widehat {dy}^2_1) + 
 F_2 F_3 (dy^2_2 + dy_3^2)  $$ 
\begin{eqnarray}
 + \  
  F_1 F_3 (dy^2_4 + dy_5^2) 
+  F_1 F_2 (dy^2_6 + dy^2_7)
+   f^{-1}dr^2+r^2d\Omega_2^2\bigg] \ , 
\label{4d11dp}
\end{eqnarray}
\begin{equation}
{\cal F}_4
 =3( *dF'^{-1}_1 \wedge dy_2\wedge dy_3  +  *dF'^{-1}_2 \wedge dy_4\wedge 
dy_5 +  *dF'^{-1}_3 \wedge dy_6\wedge dy_7 ) \ , 
\label{4d4fp}\end{equation}
where (cf. (\ref{yy}))
\begin{equation}
 \widehat {dy}_1 = dy_1 +  ({K'}\inv -1) dt  \ , 
\label{4dboost}
\end{equation}
and $K$ and ${K'}$ depend on  the  boost parameter  $\beta$ along the string 
($y_1$) direction.
The background is thus  parameterised by $f(r)$ and  the following  functions 
of $r$ ($i=1,2,3$)
\begin{equation}
K= 1+{\tilde \Q \over r}\ , \ \ \ \  {K'}\inv  = 1 - {\tilde Q \over r } K \ , 
\ \ \  \ F_{i}^{-1}=1+{\P_i \over r}\ ,\ \  \ \ {F'}_{i}\inv =1+{P_i \over r}\ ,
\label{4dfk}
\end{equation} 
\begin{equation}
 \tilde \Q =\mu \sinh^2 \beta \ , \ \  \ \ \tilde Q = \mu \sinh \beta \cosh 
\beta \ , 
\ \ \  \   \P_i = \mu \sinh^2\gamma_i \ , \ \ \ \ P_i = \mu \sh \gamma_i \cosh 
\gamma_i  \  .
\label{fffe}
\end{equation}
The four charges  $\tilde Q$ and $P_1,P_2,P_3$
are held fixed in the  limit $\mu\to 0, \beta\to \infty, \gamma_i\to \infty$.

The area of  nine-surface at  $r=\mu$   is  
\begin{equation}
A_9=  4\pi
L^7 [r^2
(K^{-1}F_1 F_2F_3)^{-1/2}]_{r=\m } =4\pi L^7\m^2 
\cosh \b  \cosh\g_1 \cosh\g_2 \cosh\g_3\ 
 ,  \label{4dareap}
\end{equation}
\begin{equation}
(A_9)_{BPS}=4\pi L^7\sqrt{\tilde Q P_1P_2P_3}\ .
\label{4dareabpsp}
\end{equation}
 The four-dimensional Einstein-frame metric  resulting upon toroidal 
compactification
is of  the form (\ref{4dst})
with
\begin{eqnarray}
\lambda (r)= ({K^{-1}F_1F_2F_3})^{1/2}
= {{r^2}\over [{
(r+\tilde \Q)
(r+\P_1)
(r+\P_2)
(r+\P_3)]^{1/2} }}\ .
\label{4dlambdap}
\end{eqnarray}
Again, this four-dimensional metric is the same 
as in  \cite{CYNex}, but  now it depends on  one
electric and  three 
magnetic  charges.

 Note that the dimensional reduction 
to ten dimensions  (along $y_1$ direction)   gives  
a  non-extreme  generalisation of a configuration
of intersecting R-R p-branes of type IIA theory,  namely, 
 a zero-brane and  
 three   four-branes \cite{CS,KTT,BL}.
Applying $T$-duality and $SL(2,Z)$ symmetry of type IIB theory one is able 
to construct various other non-extreme  $D=10$ p-brane configurations
which in the extreme limit have  a representation in terms
of intersecting $D$-branes.
Their form is always consistent with the algorithm of Section II.
In particular, it is straightforward to  write down 
the non-extreme version of the  maximally symmetric $3\bot 3\bot 3\bot
3$  solution (four three-branes intersecting at a point,  with  each pair  of
three-branes intersecting at a string), found in \cite{KTT,BL}. 

\section{$D=5$ Non-Extreme Black Holes}
The extreme   $D=5$ black holes with three independent charges
\cite{SV,TMpl} (generating solution for  general   extreme $D=5$ 
black holes with regular horizons)
can be obtained from the two different intersecting 
M-brane configurations \cite{TM}: $2\bot 2\bot2$, i.e. three
two-branes intersecting at a point,  and  ``boosted'' $2\bot 5$,
i.e. intersecting  two-brane  and five-brane with  a momentum along
the common string.
Below we shall present the  non-extreme
 versions of these  $D=11$ solutions, which serve as generating solutions for
  non-extreme  static $D=5$ black hole solutions.

\subsection{Intersection of three Two-Branes}
This   $O(4)$ symmetric 
background is a  straightforward generalisation 
of the non-extreme  $2\bot 2$  solution (\ref{mema}),(\ref{fefu})
\begin{eqnarray}
d s^2_{11} = (T_1T_2T_3)^{-1/3}
 \bigg[ - T_1T_2 T_3 f   dt^2  +  T_1 (dy^2_1 +  dy^2_2) 
+ T_2 (dy^2_3 +  dy^2_4)
 + T_3 (dy^2_5 +  dy^2_6)
\label{mm}
\end{eqnarray}
$$
 + \  f^{-1} dr^2+ r^2 d\Omega^2_{3}  \bigg],
$$
\begin{eqnarray}
{\cal F}_4 =  -3dt\wedge (dT'_1\wedge  dy_1\wedge d y_2 
+   dT'_2 \wedge  dy_3\wedge d y_4  + dT'_3 \wedge  dy_5\wedge d y_6 ) \  , 
\label{ffu}
\end{eqnarray}
where
 $$
 f= 1 - {\m \over r^2} \ , \ \ \ \ 
T^{-1}_i = 1 + {\Q_i \over r^2}\ , \ \ \   
T'_i= 1 - {Q_i \over r^2} T_i \  , \ \ \  
$$
\begin{equation}\Q_i= \m \sinh^2 \delta_i\ ,\ \ \ \ 
Q_i=\mu\sinh\delta_i\cosh\delta_i
\ , \ \ \ \  i=1,2,3\  . 
\label{mmr}
\end{equation}
For $\delta_1=\delta_2=\delta_3$, i.e.  equal $T_i$  and equal $T'_i$, 
this solution coincides with  the anisotropic six-brane solution of 
\cite{Guven}. 

The nine-area of  the regular  horizon  at $r= \mu^{1/2}$ is 
\begin{equation}
A_9 =  2\pi^2
L^6[r^3(T_1T_2T_3)^{-{1/2}}]_{r=\mu^{1/2}} 
=2\pi^2 L^6 \mu^{3/2} 
\cosh\delta_1  \cosh \d_2  \cosh \d_3\    
.  \label{5darea}
\end{equation}
In the extreme limit it becomes 
 \begin{equation}
(A_9)_{BPS}=2\pi^2 L^6\sqrt{Q_1Q_2 Q_3}\ . 
\label{5dareabps}
\end{equation}
The five-dimensional Einstein-frame  metric obtained by 
 reduction along $y_1, ...,y_6$  is 
\begin{equation}
ds^2_5=-\lambda^2(r)f(r)dt^2+\lambda^{-1}(r)\big[f^{-1}(r)dr^2+
r^2d\Omega_3^2\big]\  ,
\label{5dst}
\end{equation}
where 
\begin{equation}
\lambda (r)= (T_1T_2T_3)^{1/3}= {{r^2}\over{\left[(r^2+\Q_1)(r^2+\Q_2)(r^2+\Q_3
)
\right]^{1/ 3}}}\  .
\label{5dlambda}
\end{equation}
This  is  the  metric of non-extreme
five-dimensional black holes found in  \cite{CY5r,HMS}
(where one of the  electric charges was replaced by a magnetic one).
In the BPS limit $\Q_i \to Q_i$, $f\to 1$ and we  get  a solution which is 
$U$-dual to  the solution
of \cite{TMpl}.

\subsection{Intersection of  Two-Brane and  Five-Brane with a Boost}
The non-extreme generalisation of the supersymmetric 
configuration of a two-brane   intersecting five-brane
with a ``boost'' along the common string \cite{TM}
has the form
\begin{eqnarray}
d s^2_{11}=T^{-1/3} F^{-2/3}
\bigg[TF (- K^{-1} f dt^2 +  K \widehat {dy}^2_1) +Tdy_2^2  
+  F(dy_3^2+dy_4^2+dy^2_5 + dy_6^2)  
\label{5d11d}
\end{eqnarray}
$$
+\  f^{-1}dr^2+r^2d\Omega_3^2\ \bigg] \ , 
$$
\begin{equation}
{\cal F}_4
 =-3dt \wedge dT'\wedge dy_1\wedge dy_2 +3*dF'^{-1}_1 \wedge dy_2 \ , 
\label{5d4f}\end{equation}
where\foot{Note that $dt\wedge dy_1$ remains invariant 
under the boost, i.e. ${\cal F}_4$ does not change.}
$ \widehat {dy}_1 = dy_1 + ({K'}\inv -1) dt$.
The relevant functions of  the radial coordinate $r$ of the 
$(1+4)$-dimensional space-time are
$$ 
K = 1 + {\tilde \Q \over r^2} \ , \ \ \ 
{K'}\inv  = 1 - {\tilde Q \over r^2} K\inv \ , \ 
\ \   \tilde \Q = \m \sh^2 \b\ , \ \  \tilde Q=\mu\sinh\beta\cosh\beta
\ ,
$$
$$
T^{-1} = 1 + {\Q \over r^2}\ , \ \ \ \ \ \   {T'}= 1 - {Q \over r^2} T \  , 
\ \ \  
\Q= \m \sinh^2 \delta\ , \ \  Q=\mu\sinh\delta\cosh\delta \ ,  $$
\begin{equation}
F^{-1} = 1 + {\P \over r^2}\ , \ \ \ \ \ \ \  {F'}\inv 
= 1 + {P \over r^2}  \  , 
\ \ \  \ \ \ \ \ 
\P= \m \sinh^2 \gamma\ , \ \ P=\mu\sinh\gamma\cosh\gamma \ , 
\label{pop}
\end{equation}
and $f$ is the same as in (\ref{mmr}).  The three  charges  $\tilde Q, Q,P$
are held fixed in the extreme limit $\m\to 0, \b\to \infty, \d\to \infty, 
 \g\to \infty$.

We find again
\begin{equation}
A_9 =  2\pi^2
L^6[r^3(K^{-1} T F)^{-{1/2}}]_{r= \m^{1/2}} 
=2\pi^2 L^6 \m^{3/2} 
  \cosh\b \cosh \d  \cosh \g\ 
,  \label{5da}
\end{equation}
\begin{equation}
(A_9)_{BPS}=2\pi^2 L^6\sqrt{\tilde Q Q P}\ . 
\label{5dbps}
\end{equation}
The  corresponding five-dimensional Einstein-frame  metric 
is (\ref{5dst}) with 
\begin{equation}
\lambda (r)= (TFK^{-1})^{1/3}= {{r^2}\over{\left[
(r^2+\tilde \Q)(r^2+\Q)
(r^2+\P)\right]^{1/ 3}}} \ , 
\label{5dla}
\end{equation}
i.e.  is precisely the space-time metric found in  \cite{CY5r,HMS}.
In the extreme limit $ \tilde \Q\to 
 \tilde Q ,\  \Q \to Q, \ \P \to P $, $f\to 1$ 
 we get back to  the solution
of \cite{TMpl}.

\section{$6  \le  D\le 9$ Non-Extreme Black Holes}
  The generating solution
for black holes in dimensions $D\ge 6$  can be 
parameterised  by two charges. 
The boosted non-extreme two-brane naturally reduces to $D=9$ 
black hole.
 The $D=7$ non-extreme black hole can be  described 
as a dimensional reduction  of a configuration of two  two-branes
intersecting  at a point.
The  boosted non-extreme 
five-brane  represents the two-charge black hole in $D=6$.
Black holes in $D=10$ do not have a natural  $M$-brane description.

\subsection{Two-Brane with a Boost}
It is possible to describe all two-charge 
 $6  \le  D\le 9$ non-extreme black holes
as dimensional reductions of a non-extreme generalisation 
 of boosted two-brane 
solution which has non-maximal rotational 
isometry, i.e. $O(D-1)\times [O(2)]^{9-D}$  symmetry,  instead of $O(8)$.
Adding a boost along one of the two directions of the two-brane 
we find from (\ref{mem}),(\ref{fef})\footnote{To construct such a non-extreme 
solution
one should  start with the extreme two-brane background and assume
that the harmonic function does not depend on $D-9$ out of 9 transverse space 
coordinates, here denoted by $y_3, ...,y_{11-D}$, or, equivalently, to 
consider a periodic array of 
two-branes in these directions.}
\begin{eqnarray}
d s^2_{11} = T^{-1/3}
 \bigg[ T (- K^{-1} f  dt^2  + K \widehat {dy}^2_1 +  dy^2_2)
+ dy^2_3 + ... + dy^2_{11-D}\label{memr}
\end{eqnarray}
$$
      +\  f^{-1} dr^2+ r^2 d\Omega^2_{D-2}  \bigg] \ ,
$$
\begin{eqnarray}
{\cal F}_4 =  -3dt\wedge dT'\wedge  dy_1\wedge d y_2 \  , 
\label{fefi}
\end{eqnarray}
where $ \widehat {dy}_1 = dy_1 + ({K'}\inv -1) dt$, and
$$\  f= 1 - { \mu \over r^{D-3}} \ ,$$ 
$$ 
 K=1+ {\tilde \Q \over r^{D-3}} \ , \ \ 
 {K'}\inv =1- {\tilde Q \over r^{D-3}} K\inv  \ , 
\ \ \tilde \Q= \mu \sinh^2 \beta\ , \  \tilde Q=\mu\sinh\beta\cosh\beta
\ , $$
\begin{equation}T^{-1}  = 1 + {\Q \over r^{D-3}}\ , \ \ \   {T'} =  1 - {Q 
\over r^{D-3}} T 
\  , 
\ \  \  \ \ \ 
\Q= \mu \sinh^2 \delta \ , \  Q=\mu\sinh\delta\cosh\delta \  . 
\label{opi}
\end{equation}
$Q$ and $\tilde Q$ are the two electric  charges which are held  fixed in the 
extreme limit.

The area of the    horizon at $r=\mu^{1/(D-3)}$ is 
\begin{equation}
A_9= \omega_{D-2} L^{11-D}
[r^{D-2}
(K^{-1}T)^{-{1/ 2}}]_{r=\mu^{1/(D-3)}} =
\omega_{D-2}
 L^{11-D}
\mu^{{D-2}\over{D-3}} \ \cosh \beta \cosh\delta \ , 
 \label{Ddarea}
\end{equation}
where all  internal coordinates
$y_1, ...
,y_{11-D}$ are assumed to have period  $L$ and 
$\omega_{D-2}={{2\pi^{{D-1}\over 2}}/ {\Gamma({{D-1}\over 2})}}$.
The area (\ref{Ddarea}) { vanishes} in 
 the BPS-saturated limit, in agreement with 
 the fact  that  there are no BPS-saturated black 
holes with regular horizons of  finite   area  in  $D\ge 6$ \cite{KT,CYNear}.

The corresponding $D$-dimensional Einstein-frame metric is
\begin{equation}
ds^2_
D=-\lambda^{D-3}(r) f(r)dt^2+\lambda^{-1}(r)\big[f^{-1}(r)dr^2+r^2d\Omega_{D-2}
^
2\big]\  ,
\label{Ddst}\end{equation}
\begin{equation}
\lambda (r)= (K^{-1}T)^{{1\over {D-2}}}=  {{r^{2(D-3) \over D-2}\over
\big[(r^{D-3}+\tilde \Q)
(r^{D-3}+ \Q)
\big]^{{1\over {D-2}}}}}\ \  .\label{Ddlambda}
\end{equation} 
It coincides with the metric of  non-extreme black hole 
solutions in $D\ge 6$   \cite{Peet,HSen,CYNear}.

Note also that for $D=10$ and $\Q=0$  the  metric  (\ref{Ddst})  
describes also  the  electric 
R-R black hole  ($0$-brane) in ten dimensions \cite{HSS} 
which can be  obtained by dimensional reduction
of "boosted''  Schwarzschild solution in $D=11$.

\subsection{Five-Brane with a Boost}
The  $D=6$ black hole  with one electric and one magnetic charge
has a natural interpretation 
as a dimensional reduction of a boosted  five-brane.
Boosting the metric (\ref{fiv}) along $y_1$ we find
$$ d s^2_{11} = F^{-2/3}
 \bigg[ F \big(- K^{-1} f dt^2  + K\widehat {dy}_1^2  + dy_2^2 + ... +  dy^2_5 
 \big) $$ 
\begin{eqnarray}
+\   f^{-1} dr^2+ r^2 d\Omega^2_{4} \  \bigg]\ ,
\label{fivy}
\end{eqnarray}
where $f$ and $F$  are  the same as in (\ref{fiif}), $\widehat {dy}_1 =dy_1 +  
({K'}\inv -1)dt$  and 
\begin{equation}
K = 1 + {\tilde \Q \over r^3}\ ,
 \ \ \   {{K'}\inv } =  1 - {\tilde Q \over r^3} K\inv  \  , 
\ \ \  \ \  \tilde \Q= \m \sinh^2\beta\ , \ \ \  \tilde 
Q=\mu\sinh\beta\cosh\beta \ .
\label{fiiof}
\end{equation}
The  black hole background resulting upon  dimensional reduction 
along internal five-brane directions
$y_1,...,y_5$ is parameterised by one electric  and one magnetic charge.

\subsection{Intersection of two Two-Branes}

 The $D=7$ non-extreme black hole admits  also 
a description in terms of non-extreme 
version of $2\bot 2$ configuration.\foot{A natural description 
of two-charge $D=7$ black hole  in type IIB theory 
is given in terms of  compactified  boosted three-brane \cite{GKP}.}
Dimensional reduction of  the background 
(\ref{mema}),(\ref{fefu}) along $y_1,...,y_4$ 
leads to the  $D=7$ black hole background, with  the role
 of $\Q,\tilde \Q,Q,\tilde Q$ played by $\Q_1,\Q_2,Q_1,Q_2$.

It is also  possible to give an alternative description 
of $D=6$ black hole (now with two electric charges)
by using   $O(5)$-symmetric version 
of (\ref{mema}). i.e.   the non-extreme version of $2\bot 2$  solution 
with one of the transverse space coordinates ($y_1$) 
 treated as an isometric internal space one:
\begin{equation}
 d s^2_{11} = (T_1 T_2)^{-1/3} 
 \bigg[- T_1 T_2  f dt^2  +  dy_1^2 +  T_1  ( dy^2_2 + dy^2_3)
 +   T_2 ( dy^2_4 + dy^2_5)
\label{6d11dp}\end{equation}
$$
            + \ f^{-1} dr^2+ r^2d\Omega_{4
}^2 \  \bigg]\  .  
$$
The corresponding 
nine-area  and the  $D=6,7$ Einstein-frame 
 metrics reduce 
to  the  expressions  in (\ref{Ddarea}) and (\ref{Ddst}),
with the  role of $T$, $K^{-1}$   now played by 
$T_{1}, T_{2}$.

It is also  of interest to compare the metrics of the eleven-dimensional  
solutions 
which reduce  to black holes in  
$D=4,5,6$ with respective  $n=4,3,2$ charges  in the case when all 
charges (boost parameters)  are  equal, i.e.
$
T_i=F_i=K\inv=H\inv$. 
For $n=4$ and $n= 3$, i.e. 
 the respective cases of  $D=4$ and $D= 5$  black holes, whose 
 extreme limits have {\it  regular}
horizons,  we get 
\be
d s^2_{11} = H^{n-2}
 \bigg[- H^{-n}  f dt^2    +  f^{-1}dr^2+ r^2d\Omega^2_{6-n}  \bigg]
 +  \widehat {dy}^2_1 + ...+ dy^2_{n+3}
\ , 
\label{4dd}
\ee
where 
$\widehat {dy}_1= dy_1$ for  the unboosted configurations
(\ref{4d11d}), (\ref{mm}), 
and $\widehat {dy}_1= dy_1 + ({H'}\inv -1)dt$ 
for the  boosted ones
(\ref{4d11dp}), (\ref{5d11d}). 
At the same time, in the case of $n=2$, i.e. black holes
whose extreme limits  have singular horizons, e.g., for   $D=6$ black hole, 
 we get the metric 
\be
d s^2_{11} = H^{2/3}
 \bigg[ H\inv  (-  f dt^2  + dy^2_2 + ...+ dy^2_5) +  \widehat {dy}^2_1 
 +  f^{-1}dr^2+ r^2d\Omega^2_{4}  \bigg]\ , 
\label{ggg}
\ee
where  $\widehat {dy}_1 = d y_1$  for the  intersecting two-brane 
representation 
(\ref{6d11dp}) and $\widehat {dy}_1= dy_1  +  ({H'}\inv -1)dt$ 
for the boosted five-brane one  (\ref{fivy}).
 While the radii of the internal coordinates are  constant 
in the case  (\ref{4dd}), i.e.  $D=4,5$ 
 black holes with equal $n=3,4$ charges,  this is no
longer  so for  black holes  in  $D>5$ with equal $n=2$ charges.\footnote{Note that 
``bound-state'' metrics  (\ref{4dd}),(\ref{ggg}) are  different from
 the isotropic 
black p-brane ones  discussed  in  \ci{DLP,KT}.}

\section{Universal Expressions for Mass and Entropy}\label{Conclusions}
It is possible to write down  the  formulas for the mass 
and the entropy  which apply to  all eleven-dimensional  ``anisotropic
p-brane''  solutions discussed in previous Sections.  Similar
expressions for the corresponding non-extreme black holes were given
in \cite{CYNex,CYNear}. Note also that analogous relations for 
isotropic black brane solutions  appeared in  \ci{DLP,KT}.

Let $p=11-D$ be the  common internal dimension of intersecting
M-branes, i.e.
 the dimension of an  anisotropic p-brane. 
 Then  $D$ 
is the dimension of the  black hole obtained by dimensional reduction.  
For $D=4$, $D=5$ and $6\le D\le 9$   the black hole has the metric of the form 
 (\ref{Ddst})
 with  $\l (r)=  (H_1 ...H_n)^{-1/(D-2)}$, \ $H_i= 1 +
{\Q_i}/r^{D-3}$,
\ $i=1,...,n$,\   with $n\le4$, $n\le 3$ and $n\le 2$, respectively
(cf. (\ref{4dlambda}),(\ref{4dlambdap}),
(\ref{5dlambda}),(\ref{5dla}),(\ref{Ddlambda})). 
Here we  shall use the same   notation $\Q_i$ for all $n$ charges, some of 
which may be electric, and some  magnetic.
Let $G_{11}=8\pi \k^2$ be the Newton's constant in  eleven dimensions
(the Newton's constant in $D$ dimensions  is then
$G_{D}= G_{11}/L^p$).
Then  for a $O(D-1)$ - symmetric solution
 the  normalised charges per unit  volume  are\footnote{The relation
of our notation to that of 
\ci{KT} is the following: $D-3\to d $, \ $ Q \to r_0^d$, \ $\Q \to r^d_-$, \ 
$\m \to \m^d$.  The  charges  $q_i$ correspond to constituent 
objects of an $N$-charge bound state (hence there is  no $\sqrt n$
factor
 in the  expression for the charges).}
\be
q_i =  a  Q_i \ , \ \ \ \ a\equiv {\omega_{D-2}\ov \sqrt 2  \k }  (D-3)   
\ , \ \ \  \ \ \  
Q_i =  \m \sh \b_i \ch \b_i\ , \ \ 
  \Q_i=\mu\sinh^2\beta_i\ . 
\la{uyru}
\ee
 As  follows from (\ref{Ddlambda}), the ADM mass  is 
\be
 M_{ADM}  = {\omega_{D-2}\ov 2\k^2 } L^p  \  [ (D-2) \m +
  (D-3) \sum^n_{i=1} \Q_i ] \ .
\la{mas}
\ee
It  can  be  expressed in terms of  the non-extremality
parameter
$\m$ and charges $Q_i
$  (which are fixed in the extreme limit $\mu\to 0$)
as  follows 
 (note that  $\Q_i =  - {\m\ov 2} + \sqrt {Q_i^2 + ({\m\ov 2})^2} $) 
\be
 M_{ADM} =  b \ \bigg[ \sum^n_{i=1} 
\sqrt {Q_i^2 + ({\m\ov 2})^2}  + \l  \m  \bigg] \ ,  \ \ \ \ \ \ \ b \equiv 
{\omega_{D-2}\ov 2\k^2 } (D-3) L^p \ , 
\la{mesu}
\ee
where the
 parameter  $\l$ (not to be confused with the function $\l(r)$ in
 $D$-dimensional  metric (\ref{Ddst})) is 
the same as in \ci{KT} 
\be 
 \l  \equiv {D-2\ov D-3  } -  {n\over 2}  \ .   
\la{lll}
\ee
Explicitly,   $\l_{n=1} =  {D-1\ov 2(D-3 ) },
 \ \l_{n=2} =  {1\ov D-3  }, \ \l_{n=3} =  {5-D\ov 2(D-3)  }, \ \l_{n=4} =  
{4-D\ov 2(D-3 ) }$, i.e.  $\l \ge 0$  and 
  {\it vanishes } only for  $D=4,\  n=4$ and $D=5,\  n=3$,
i.e.  the cases with  regular horizons
 in the extreme (BPS-saturated)   limit.\foot{The above
 expressions for  the  charges and the  mass include
 all previously discussed non-extreme cases: 
 single  two-brane ($p=2$, $D=9$), single   five-brane
($p=5, D=6$), 
and equal-charge  $p=4$ ($D=7$)  and $p=6$ ($D=5$) 
 anisotropic branes \ci{Guven}\ . Note that  our expression  for
 the mass in $p=4$ case disagrees with that 
in \ci{Guven}.}

The  Bekenstein-Hawking  entropy $S_{BH}$, 
which follows from the obvious generalisation of the 
expressions for the area  (\ref{Ddarea}),
 (\ref{4darea}),(\ref{4dareap}),(\ref{5darea}) and (\ref{5da}) (with
$K\inv T \to (H_1 ...H_n)\inv $)
is 
\be
S_{BH}= {2\pi A_9\ov \k^2} = c \m^{{D-2}\over{D-3}} \prod_{i=1}^n \ch \d_i 
\ ,  \ \ \ \ \ \  \ \ c\equiv {2\pi\omega_{D-2} \ov \k^2} L^{p} \ , 
\la{arrq}
\ee
or, in terms of the Hawking temperature $T_H$, 
\be
S_{BH} =  b \  {  \m  \ov   T_H } \ , 
\ \ \ \ \ \ \ \ 
\      T_H = {1\ov 4\pi } (D-3) \m^{-{1\ov D-3}}  \prod_{i=1}^n (\ch \d_i)\inv 
 \ .   
\la{awq}
\ee
These general formulas apply also  to the  case of $D=10$ black hole \ci{HSS}
where $D=10$ and $n=1$. 

Expressed in terms of $\m$ and $Q_i$, the entropy becomes
\be
S_{BH}=   c \m^\l  \prod_{i=1}^n \bigg[  \sqrt {Q_i^2 + ({\m\ov 2})^2 } + { \m 
\ov 2}  \bigg]^{1/2} \ .  
\la{arroq}
\ee
It  has  non-zero  extreme limit ($\m\to 0$)  only 
when  $\l=0$. In this case $M_{ADM}$ and $S_{BH}$ take 
simple forms
\be 
M_{ADM} =  b  \sum^n_{i=1} 
\sqrt {Q_i^2 + ({\m\ov 2})^2}  \ , 
\ \ \ \ \ \  S_{BH}=   c \prod_{i=1}^n \bigg[  \sqrt {Q_i^2 + ({\m\ov 2})^2}  
+ { \m \ov 2} \bigg]^{1/2} \ .  
\la{wrq}
\ee
 $M_{ADM}$ resembles the  energy of a system of relativistic particles 
with masses $Q_i$ (masses of individual constituents in extreme limit), 
all having  the same  momentum proportional to 
 $\m$. This  suggests  a  ``bound-state'' interpretation of 
 this non-extreme  system 
 (cf. \ci{DR}).

For all  fixed values of $\l$ and 
$Q_i$ the mass (\ref{mesu})  and  the entropy (\ref{arroq}) satisfy the following relation 
\be
{  \partial  \ln  S_{BH} \over  \partial \ln  \mu}   = b^{-1}  {    \partial M_{ADM} \over  \partial \mu } \ , \ \ \ 
i.e. \ \ \ \ 
{  T_H} {  \partial  S_{BH} \over  \partial  \mu}   =     {    \partial M_{ADM} \over  \partial \mu } \ . 
\la{poy}
\ee
This 
 thermodynamic relation 
 is  valid both  in the Schwarzschild  ($Q_i=0$) case  as well as in 
the extreme limit ($\m=0$). 
This explains why 
the proportionality  between the entropy and the area
of the horizon can be assumed
to be true also in the extreme limit
(cf. \ci{Horowitz}): this proportionality 
 certainly holds  in non-extreme (or near-extreme) case (see, e.g., \ci{HH})
and thus should be meaningful
 also in the limit  $\m \to 0$.

Other intersecting M-brane configurations (with $\lambda \ne 0$) have
zero entropy in the  extreme limit ($\mu\to 0$).
Representative examples 
of such configurations  are   unboosted  $p=7$
configurations $5\bot5\bot5$, $5\bot5\bot2$, $5\bot 2\bot 2$,
corresponding to $D=4$ black holes with $n=3$ charges,  
 unboosted $p=6$ configuration $2\bot 5$, corresponding to $D=5$ black
hole with $n=2$ charges, and  $p=4$ 
configuration $2\bot 2$, corresponding to $D=7$ black hole with $n=2$
charges. In this case  it is of interest to study the near-extreme  limit,
 where
\be 
M_{ADM} = M_0 + \Delta M +  O(\m^2) \ , \ \ \ \ \  M_0 =  b  \sum^n_{i=1} 
Q_i   \ , \ \ \ \  \Delta M=  b \l \m      \ , 
\la{oop}
\ee
\be
 S_{BH}= c_1 \prod_{i=1}^n  { Q_i^{1/2} }\  E^\l  \  , \
 \ \ \  E\equiv \Delta M  \ , \ \ \ \  \  c_1= c(b\l)^{-\l} \ . 
\la{wrqe}
\ee 
 Using  the   thermodynamic relation
$dE= T dS$ it follows from (\ref{wrqe}) that 
\be
 S_{BH}= c_2 \prod_{i=1}^n ({Q_i})^{1\ov 2(1-\l)} \  T^{\l\ov 1-\l}   \  , \ \ 
\ \
\ \ \ c_2 = (c_1 \l^\l)^{1\ov 1-\l} = (c b^{-\l})^{1\ov 1-\l} \ , 
\la{iii}
\ee
where 
\be
T =  (c_1 \l)\inv   \prod_{i=1}^n {Q_i}^{-1/2} \  E^{1-\l}\ 
\la{wew}
\ee
is  the near-extreme limit of the Hawking temperature $T_H$ in 
(\ref{awq}).

Generalising the discussion in \ci{KT},  we may enquire when this entropy has a 
massless ideal gas entropy form.
The power  $\n ={\l/(1-\l)}$
 of the temperature  in  (\ref{iii})   is  equal to 2 and 5   for 
 unboosted two-brane and five-brane, respectively  \ci{KT}.\footnote{In particular, for the five-brane ($D=6,  n=1,$ $ \  \l= {5\ov 6},\  \n =5 $)
eq. (\ref{iii}) gives the expression found in 
\ci{KT} $ S_{BH} = 2^7 3^{-6} \pi^3 n^3 L^5 T^5 $ 
after one uses the charge  quantisation condition
$q= {3\omega_4  \ov \sqrt 2 \k} Q = ({\sqrt 2 \pi \ov \k})^{1/3} n $.}
The only other cases when $\n$ is integer   
are  configurations with $D=5,\  n=2$,  i.e. the anisotropic six-brane 
corresponding to  unboosted $2\bot 5$ intersection, 
and $D=4,\  n=3$, i.e.  the anisotropic seven-branes corresponding to  
 unboosted  $5\bot 5\bot 5$, 
$2\bot 5\bot 5$ and $2\bot 2\bot 5$ intersections.
In these cases  $\l=1/2$ and  $\n=1$, so that,  not unexpectedly \ci{KTT}, 
here   $S_{BH}$ has  a  ``string''-like form, i.e. 
the form  of   the entropy of 
a gas  of massless particles in (1+1)-dimensions.
For other anisotropic  eleven-dimensional p-branes,  e.g., those reducing to   
 isotropic dilatonic p-branes in lower dimensions 
\ci{DLP,KT},  $S_{BH}$ does not have an ideal gas scaling.

To conclude, we
have constructed  non-extreme versions of
 intersecting M-brane solutions which correspond to  one parameter
``deformations'' of  the  supersymmetric intersecting M-brane
solutions, and  maintain the  simple ``product'' structure.
  This 
 product structure  implies that the non-extreme static black holes
obtained upon dimensional reduction  have a form   which provides  a
 straightforward interpolation  between  the Schwarzschild and 
BPS-saturated  backgrounds.

The M-brane interpretation of non-extreme
black hole solutions may provide  an insight into the problem of statistical 
understanding  of their properties.
In particular,  it would be  of interest to  study in more detail 
 the statistical origin of the  BH entropy of 
 near-extreme  anisotropic p-branes  and interpret  them 
in terms of  massless modes living on 
near-extreme  intersections,   along the lines of  \ci{KT,KTT}.


\acknowledgments
We  would like to thank  I. Klebanov, G. Papadopoulos, J. Russo 
 and P. Townsend for useful 
discussions and  remarks. M.C.  acknowledges the  hospitality of the 
Department 
 of Applied Mathematics and Theoretical Physics  of  Cambridge 
University. 
The work is supported by U.S. DOE Grant No. DOE-EY-76-02-3071 (M.C.),
 the National 
Science Foundation Career Advancement Award No. PHY95-12732 (M.C.), the 
PPARC (A.T.), ECC grant SC1$^*$-CT92-078 (A.T.) 
and  the NATO collaborative research grant CGR No. 940870.

\vskip2.mm


\end{document}